# Wave chirality reversal without passing achirality

**M.V. Berry[1,*], K.Y. Bliokh[2,3] & J.M. Robbins[4]**

[1]H H Wills Physics laboratory, Tyndall Avenue, Bristol BS8 1TL, UK

[2] Donostia International Physics Center (DIPC), Donostia-San Sebastián 20018, Spain

[3] IKERBASQUE, Basque Foundation for Science, Bilbao 48009, Spain

[4] Fry Building, Woodland Road, Bristol BS8 1UG, UK

[1]asymptotico@bristol.ac.uk, [2]konstantin.bliokh@dipc.org, [4] j.robbins@bristol.ac.uk

## Abstract

Chiral connectedness describes chiral structures that transform smoothly into their mirror images while never being achiral. The phenomenon is illustrated with a two-parameter family of complex vector fields $\boldsymbol{E}$ representing paraxial optical Gaussian beams. $\boldsymbol{E}$ is chiral everywhere in the parameter plane except the origin. During a circuit of the origin, $\boldsymbol{E}$ reverses chirality twice without being achiral. Different aspects of chirality are associated with the vector and scalar nature of $\boldsymbol{E}$; natural measures of each, and their combinations, change sign at parameter values where $\boldsymbol{E}$ is chiral. The analysis is simpler for planar fields $\boldsymbol{E}$, but their chiralities are essentially the same when the subtleties of three dimensions are included.

Orcid IDs: M V Berry 0000-0001-7921-2468, K Y Bliokh 0000-0003-1104-8627, J M Robbins 0000-000108772-4316

*Author to whom correspondence should be addressed

## 1. Introduction

It is known that chiral objects can be transformed into their mirror images without passing through an achiral shape: 'chiral connectedness' [1, 2]. A rubber glove can be transformed from left- to right-handed by turning it inside-out finger by finger. If the object consists of $n$ point particles in $d$ dimensions, chiral connectedness requires $n \geq d + 2$ [3-6].

This has the important implication that although an object can be identified as being chiral or not, there can be no purely geometric consistent assignment of handedness; the concepts of chirality and handedness are distinct [2]. Any single scalar measure of chirality (e.g.[7]), whose sign purports to specify handedness, must pass through zero, and chiral connectedness means that at such a 'blind spot' [1] the object can nevertheless be chiral. In practice, assignments of chirality are relative to how an object is interrogated. A child can sort a pile of gloves into left- and right-handed by comparison with her hands (the inside-out transformation is disregarded), but this sorting fails for potatoes which are also generally chiral; and for interrogation by rotation upon scattering of linearly polarised light [8], zero polarisation rotation by a chiral object is usually not considered (or not important).

Chiral connectedness suggests that instead of attempting to quantify handedness with a simple measure, shapes of objects are specified by parameters, and paths considered in the parameterised space of shapes. Our modest contribution to this approach concerns not continuous objects or collections of discrete points, but a particular two-parameter family of complex vector fields, representing for example paraxial light beams. For such fields, there are two kinds of chirality: associated with the vector nature of the field (in optics, related to spin angular momentum, representing polarisation); and associated with the complex scalar nature of the field (in optics, related to orbital angular momentum). Simple measures of chirality can be defined, and subsequently combined to give new measures of the field's chirality. But whatever the prescription, fields for which any such measure vanishes are nevertheless typically chiral.

With parameters $a, b$, our family is chosen to be chiral everywhere in the $a, b$ plane except the origin, where it is achiral. During a circuit of the origin, the chirality reverses twice, so our field illustrates chiral connectedness.

To expose chiral connectedness in the simplest way, we first consider complex vector fields in the plane. In section 2, the spin and orbital chiralities are defined for general complex planar vector fields. Section 3 describes our particular family of fields, in which spin and orbital aspects are entangled, and discusses the two chiralities.

When our vector field is regarded as inhabiting three-dimensional space, more subtle considerations enter, as described in section 4. Nevertheless, the chirality measures are essentially the same as in two dimensions. Section 5 contains some concluding remarks.

## 2. Polarisation and orbital chiralities in the plane

We consider a complex vector field $\boldsymbol{E}(\boldsymbol{r})$ in the plane $\boldsymbol{r} = (x, y)$. In the Cartesian basis,

$$\boldsymbol{E}(\boldsymbol{r}) = \left(E_x(\boldsymbol{r}), E_y(\boldsymbol{r})\right). \tag{2.1}$$

It is convenient to work with normalised fields, and introduce local densities as well as integral values, denoted by angle brackets, i.e.

$$\langle\, i(\boldsymbol{r})\rangle \equiv \iint d\boldsymbol{r}\; i(\boldsymbol{r}) = 1, \tag{2.2}$$

where the integral is over the $\boldsymbol{r}$ plane, and the local intensity is

$$i(\boldsymbol{r}) = |\boldsymbol{E}(\boldsymbol{r})|^2. \tag{2.3}$$

If $\boldsymbol{E}(\boldsymbol{r})$ represents the electric field of a wave, its vector nature is associated with polarisation (spin) angular momentum, and its scalar nature (phase and intensity) is associated with orbital angular momentum. Each is associated with a different aspect of chirality [9].

For spin, it is convenient to use the circular basis:

$$E_{\pm}(\boldsymbol{r}) = \frac{1}{\sqrt{2}}\left(E_x(\boldsymbol{r}) \mp iE_y(\boldsymbol{r})\right). \tag{2.4}$$

Then a natural measure of spin chirality is the difference of the left- and right-handed intensities [10], equivalent to the total spin $\langle S_z \rangle$:

$$C_{sp} = \langle\, c_{sp}(\boldsymbol{r})\rangle\,, \tag{2.5}$$

in which

$$c_{sp}(\boldsymbol{r}) = |E_+(\boldsymbol{r})|^2 - |E_-(\boldsymbol{r})|^2 = 2\mathrm{Im}\left[E_x^*(\boldsymbol{r})E_y\,(\boldsymbol{r})\right] = S_z(\boldsymbol{r}). \tag{2.6}$$

For orbital, we use the basis of angular-momentum eigenstates; in polar coordinates, this is

$$\boldsymbol{E}(r) = \sum_{l=-\infty}^{\infty} \boldsymbol{E}_l(r)\exp(il\phi). \tag{2.7}$$

It is natural to define the orbital contribution to the chirality as the $l$-weighted difference between the $l > 0$ and $l < 0$ intensities, equivalent to the total orbital angular momentum $\langle L_z \rangle$:

$$C_{orb} = \langle c_{orb}(\boldsymbol{r})\rangle, \tag{2.8}$$

in which

$$\begin{aligned} c_{orb}(\boldsymbol{r}) &= r\sum_{l=-\infty}^{\infty} l|\boldsymbol{E}_l(r)|^2 = r\sum_{1}^{\infty} l(|\boldsymbol{E}_l(r)|^2 - |\boldsymbol{E}_{-l}(r)|^2) \\ &= \boldsymbol{E}^*(\boldsymbol{r})\cdot\left(-ix\partial_y + iy\partial_x\right)\boldsymbol{E}(\boldsymbol{r}). \end{aligned} \tag{2.9}$$

We note that the local function $c_{orb}(\boldsymbol{r})$ is extrinsic, because it depends on the choice of origin of $\boldsymbol{r}$. However, the integrated quantity $C_{orb}$ , like $C_{sp}$, is a quantity intrinsic to the field, i.e. independent of choice of origin, provided the total in-plane momentum vanishes: $\langle \boldsymbol{p} \rangle = \langle \mathrm{Im}[\boldsymbol{E}(\boldsymbol{r})^* \cdot (\nabla)\boldsymbol{E}(\boldsymbol{r})] \rangle = \boldsymbol{0}$ [11].

Scalar measures of the chirality of the field $\boldsymbol{E}(\boldsymbol{r})$ can be defined as $C_{sp}$ or $C_{orb}$ separately, or combinations such as the two simple spin-orbital democratic choices

$$C_1 = C_{sp} + C_{orb}\ , \ C_2 = C_{sp}C_{orb}\ . \tag{2.10}$$

In practice, measures are selected that depend on the physics of how chirality is detected. We will see that different measures vanish for different fields $\boldsymbol{E}(\boldsymbol{r})$ that are nevertheless chiral - blind spots reflecting chiral connectedness.

## 3. Optical example

A two-parameter family of normalised complex vector fields that illustrates chiral connectedness is

$$\boldsymbol{E}(\boldsymbol{r}; a, b) = \frac{\sqrt{8}}{\sqrt{\pi(1+a^2)(1+b^2)}} \exp(-x^2 - y^2)\,(x + iby)\big(\boldsymbol{e_x} + ia\boldsymbol{e_y}\big). \tag{3.1}$$

The parameters are $a$ and $b$, both real. This represents any transverse section of an entangled superposition of Hermite-Gauss beams ($HG_{01}$ and $HG_{10}$). For all parameters in the $a, b$ plane, except the origin $a = b = 0$, this field is chiral. (With the parameters $a, b$ regarded as complex, the class of complex vector fields (3.1) is richer, but in the following we will not need this generality.)

Any fields (3.1) at antipodal points $(a, b)$ and $(-a, -b)$ are mirror images. Therefore any path connecting a pair without passing through the origin accomplishes chirality-reversal without encountering achirality, i.e. chiral connectedness. This is illustrated in figure 1.

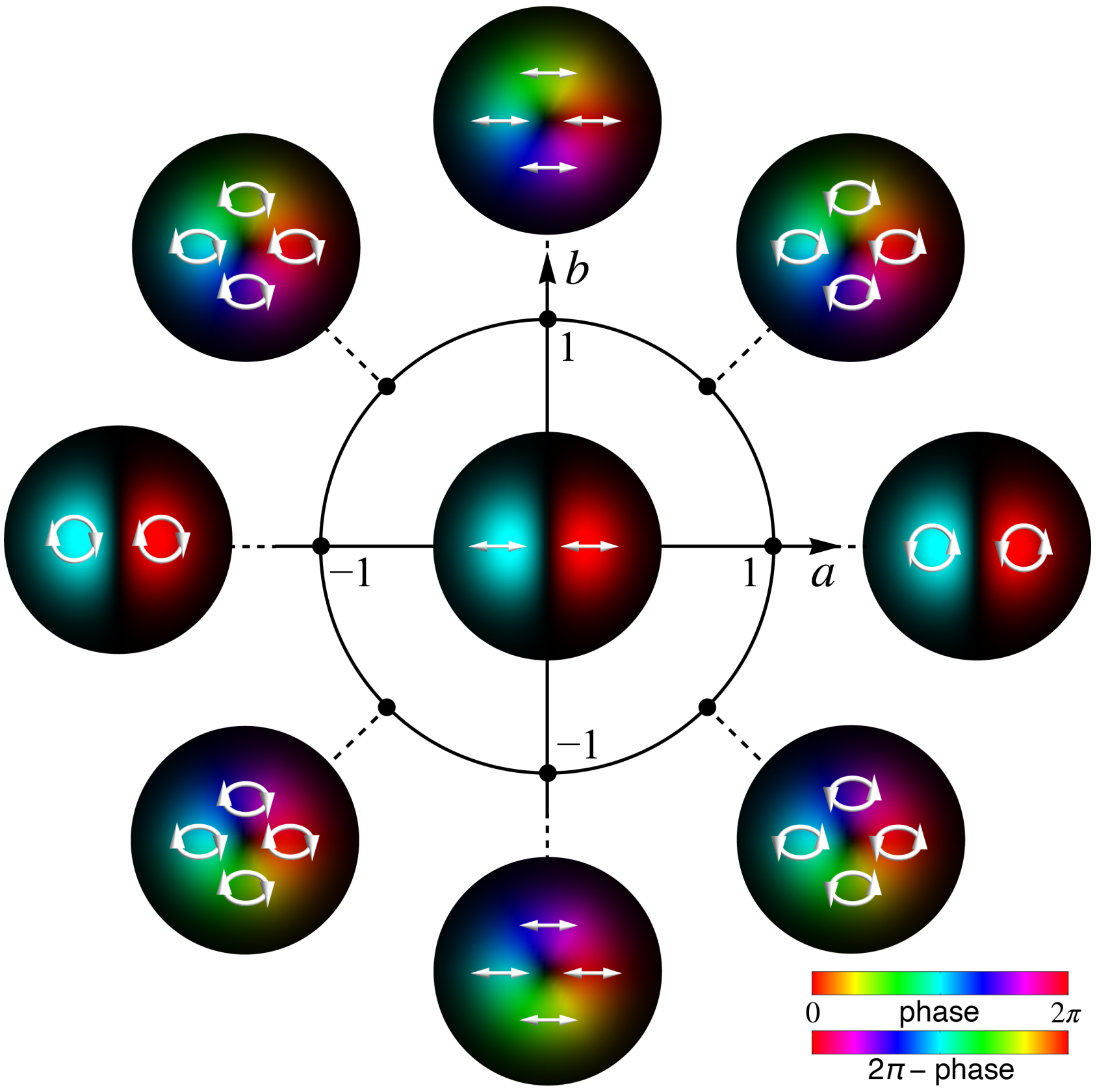

Figure 1. Depiction of the field (3.1) around the unit circle in the $(a, b)$ plane enclosing the origin. The white loops represent the polarisation ellipses associated with $\boldsymbol{e}_x + ia\boldsymbol{e}_y$; the colours represent the phase $\arg(x + iby)$, coded by hue in the range 0 (red) to $2\pi$ (red), with brightness denoting the amplitude $|\boldsymbol{E}|$; the coloured bars indicate phase and the mirror-image $2\pi$ −phase. At the achiral origin (0,0), cyan corresponds to phase $\pi$, and red to phase 0 $(2\pi)$ ; both are fixed points of the map phase→ $2\pi$ −phase.

It is straightforward to calculate the local densities (2.3), (2.6)) and (2.9), and the chiral measures (2.5) and (2.8). The densities are

$$
\begin{aligned}
i(\boldsymbol{r}) &= \frac{8\exp\big(2(x^2+y^2)\big)}{\pi(1+b^2)}(x^2+b^2y^2) \\
c_{sp}(\boldsymbol{r}) &= \frac{16a\exp\big(2(x^2+y^2)\big)\,(x^2+b^2y^2)}{\pi(1+a^2)(1+b^2)} \\
c_{orb}(\boldsymbol{r}) &= \frac{8\exp\big(2(x^2+y^2)\big)}{\pi(1+b^2)}(x^2+y^2).
\end{aligned}
\tag{3.2}
$$

Note that $i(\boldsymbol{r})$ and $c_{sp}(\boldsymbol{r})$ are identical apart from a prefactor, and $c_{orb}(\boldsymbol{r})$ (where the origin of $\boldsymbol{r}$ has been chosen as the Gaussian axis) possesses rotational symmetry.

The spin and orbital chiralities are easily calculated from (2.5) and (2.8):

$$
C_{sp}(a,b) = \frac{2a}{1+a^2}, \tag{3.3}
$$

and

$$
C_{orb}(a,b) = \frac{2b}{1+b^2}. \tag{3.4}
$$

(We note that for the orbital chirality, in the representation (2.7), the only nonzero angular momentum components are $\boldsymbol{E}_{\pm 1}$.)

Now the two chirality measures (2.10) are

$$
C_1(a,b) = \frac{4(a+b)(1+ab)}{(1+a^2)(1+b^2)},\; C_2(a,b) = \frac{4ab}{(1+a^2)(1+b^2)} \tag{3.5}
$$

Each of the four measures vanishes at ‘blind spots’ that are chiral; as figure 2 illustrates, the blind spots of each of $C_{sp}$ and $C_{orb}$ coincide with some of those of $C_2$, but are separate from those of $C_1$, explicitly illustrating the inadequacy of any single chirality measure [1]

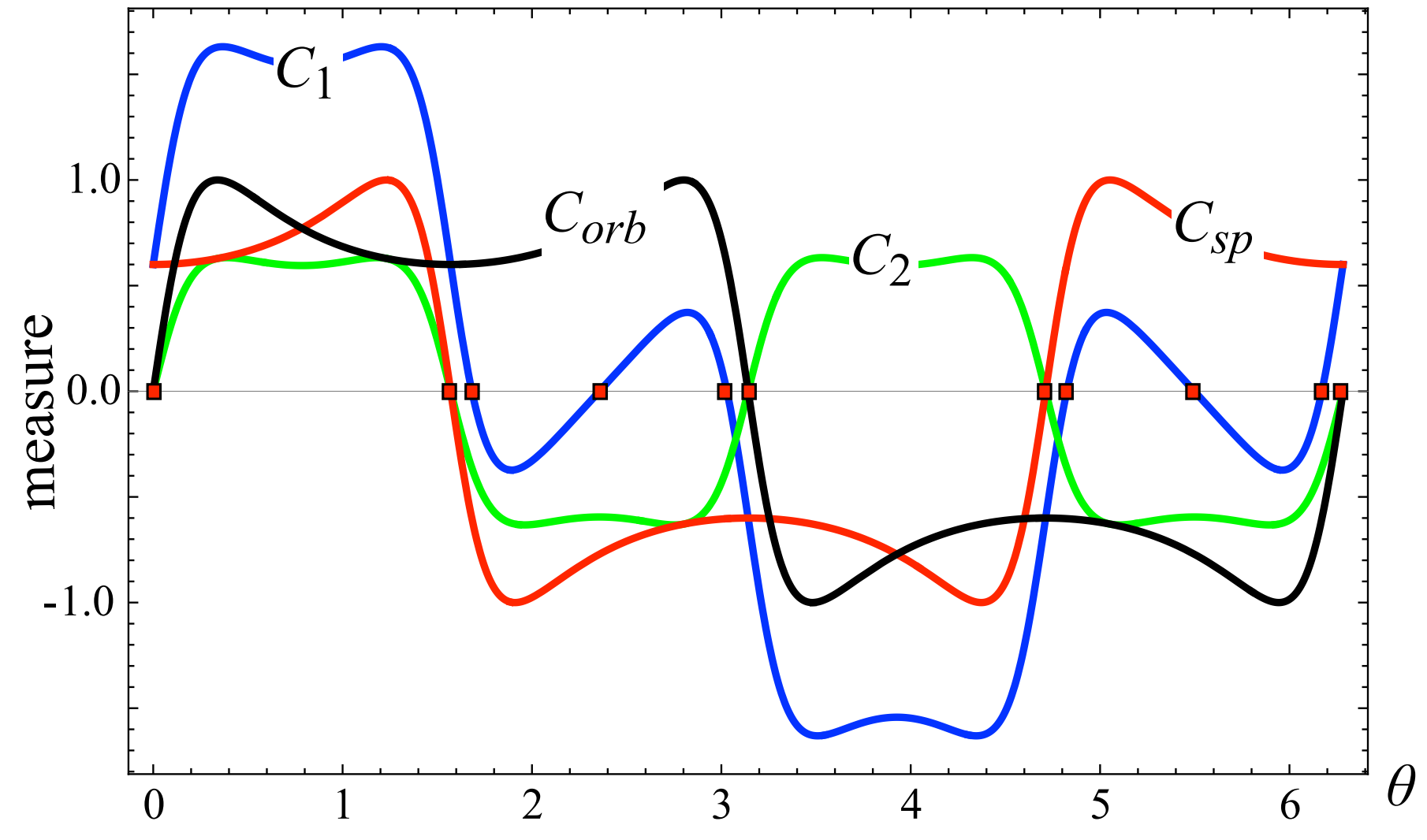

Figure 2. The four chirality measures $C_{sp}, C_{orb}, C_1, C_2$ on a circuit $a = \rho\cos\theta,\ \ b = \rho\sin\theta$, for $\rho = 3$ and $0 \leq \theta \leq 2\pi$

For any two (or more)-parameter family of complex vector fields, an interesting question is: can the field acquire a geometric phase during a cycle $\mathcal{C}$ of the parameters? For the family (3.1), the answer is: no. To show this, we start from the general formula for the phase [12], as the integral around $\mathcal{C}$ of a 1-form $\boldsymbol{A}(a, b)$:

$$\gamma(\mathcal{C}) = \oint d(a, b) \cdot \boldsymbol{A}(a, b), \tag{3.6}$$

in which

$$\boldsymbol{A}(a, b) = \iint d\boldsymbol{r}\ \ \mathrm{Im}\big[\boldsymbol{E}^*(\boldsymbol{r}; a, b) \cdot \big(\nabla_{(\boldsymbol{a},\boldsymbol{b})}\big)\boldsymbol{E}(\boldsymbol{r}; a, b)\big]. \tag{3.7}$$

From (3.1), the '1-form density' is

$$\mathrm{Im}\big[\boldsymbol{E}^*(\boldsymbol{r}; a, b) \cdot \big(\nabla_{(a,b)}\big)\boldsymbol{E}(\boldsymbol{r}; a, b)\big] = \left(0, \frac{8\exp\big(-2(x^2 + y^2)\big)}{\pi(1 + b^2)} xy\right). \tag{3.8}$$

This is odd in $\boldsymbol{r}$, so the 1-form vanishes:

$$\boldsymbol{A}(a, b) = 0, \tag{3.9}$$

and $\gamma(C) = 0$. This result can also be understood from the fact that the 1-form density (3.8) depends only on $b$, so no area is enclosed while cyclically varying parameters in the mode (3.1); this is true in the $a, b$ plane as well as on the polarisation Poincaré sphere and its orbital analogue [13]).

**4. From two dimensions to three**

Any purely planar field , e.g. $\boldsymbol{E}(\boldsymbol{r}; a, b)$, can be regarded as embedded in 3d, with a $z$ direction perpendicular to the $\boldsymbol{r}$ plane. Then its 2d chirality depends on which side it is viewed from; in other words, the original chirality can be reversed by a $\pi$ rotation including $z$, for example about the $y$ axis. In this sense all purely 2d objects or fields are achiral in 3d. But for the planar field (3.1) considered physically, as a section through a paraxial optical beam in 3d, with propagation along $z$ embodied in the additional phase factor $\exp(ikz)$, where $k$ is the propagation wavenumber, the field is not purely 2d, and can be 3d-chiral. Alternatively stated, a natural choice of viewing is along the propagation direction: along $+z$ for $k > 0$ and $-z$ for $k < 0$. This introduces a factor $\mathrm{sgn}(k)$ into 3d chiralities.

To demonstrate chiral connectedness in 3d, we note that a chiral image of the initial field is provided by the parity (spatial inversion) transformation P: $(\boldsymbol{r}, z) \to (-\boldsymbol{r}, -z)$, which transforms the field (3.1), including the phase factor $\exp(ikz)$, as $\boldsymbol{E}(\boldsymbol{r}; a, b)\exp(ikz) \to \boldsymbol{E}(\boldsymbol{r}; a, b)\exp(-ikz)$. To continuously perform this transformation, start with $\boldsymbol{E}(\boldsymbol{r}; a, b)\exp(ikz)$ and go around a half-circle in the $a, b$ plane, to obtain $\boldsymbol{E}(\boldsymbol{r}; -a, -b)\exp(ikz)$. Then perform a spatial $\pi$ rotation about the $y$ axis, reversing $z, x, \boldsymbol{e}_x$, to obtain $\boldsymbol{E}(\boldsymbol{r}; a, b)\exp(-ikz)$, which is the chiral image of the initial field. During this process, the field has never been achiral, but any chirality measure (e.g. the spatial integrals of (4.1) and (4.2) to follow) must pass through zero.

From a wider perspective [8, 14], the 3d chirality of a dynamical or wave system should be quantified by a pseudo-scalar quantity that changes sign under parity transformation P but remains invariant under time reversal (T). For the vector (polarisation) degrees of freedom of a monochromatic field, this pseudoscalar is the helicity: the scalar product of the field with its curl [15-19]), or, for paraxial fields, the projection of spin onto the propagation direction. For the optical field $\tilde{\boldsymbol{E}}(\tilde{\boldsymbol{r}}; a, b) = \boldsymbol{E}(\boldsymbol{r}; a, b)\exp(ikz)$, with $\tilde{\boldsymbol{r}} = (x, y, z)$ and the corresponding gradient operator $\tilde{\nabla}$, the helicity density can then be written as

$$\sigma(\boldsymbol{r}) = \frac{1}{|k|}\mathrm{Re}\big[\tilde{\boldsymbol{E}}^* \cdot \big(\tilde{\nabla} \times \tilde{\boldsymbol{E}}\big)\big] = \boldsymbol{S} \cdot \frac{\boldsymbol{k}}{|k|} = c_{sp}(\boldsymbol{r})\,\mathrm{sgn}(k), \tag{4.1}$$

where $\boldsymbol{S} = \mathrm{Im}(\boldsymbol{E}^* \times \boldsymbol{E})$ is the spin angular momentum density [11, 20, 21] and $\boldsymbol{k} = k\,\boldsymbol{e}_z$ is the wavevector. Note that $\tilde{\nabla} \times \tilde{\boldsymbol{E}}$ is proportional to the magnetic field, so $\sigma(\boldsymbol{r})$ embodies 'electric-magnetic democracy' [22], and the magnetic chirality need not be considered separately.

We note that the helicity or spin chirality density admits a well-defined description by (4.1) via polarisation of the field at a given point. In contrast, the orbital chirality density, which is associated with the spatial geometry of the field, does not admit a purely local definition: the projection of the local 3d orbital angular momentum density $\boldsymbol{L}(\boldsymbol{r}) = \tilde{\boldsymbol{E}}^* \cdot \big(-i\boldsymbol{r} \times \tilde{\nabla}\big)\tilde{\boldsymbol{E}} = \boldsymbol{r} \times \tilde{\boldsymbol{p}}(\boldsymbol{r})$ (where both longitudinal and transverse components of $\tilde{\boldsymbol{p}}(\boldsymbol{r})$ and $\boldsymbol{L}(\boldsymbol{r})$ are important) onto the local momentum density $\tilde{\boldsymbol{p}}(\boldsymbol{r})$ vanishes. Instead, the orbital chirality density can be defined as the projection of $\boldsymbol{L}(\boldsymbol{r})$ onto the direction of the integrated momentum, which is purely longitudinal for the field under consideration, $\langle\tilde{\boldsymbol{p}}\rangle/|\langle\tilde{\boldsymbol{p}}\rangle| = \mathrm{sgn}(k)\,\boldsymbol{e}_z$:

$$\ell(\boldsymbol{r}) = \boldsymbol{L}(\boldsymbol{r}) \cdot \frac{\langle\tilde{\boldsymbol{p}}\rangle}{|\langle\tilde{\boldsymbol{p}}\rangle|} = c_{orb}(\boldsymbol{r})\,\mathrm{sgn}(k). \tag{4.2}$$

Equations (4.1) and (4.2) have the desired forms of T-symmetric pseudoscalar densities in 3d (as angular momentum is P-even and T-odd, whereas momentum is P-odd and T-odd). Integrating (4.1) and (4.2) yields the 2d chiralities (3.3) and (3.4) with $\mathrm{sgn}(k)$ factors.

## 5. Concluding remarks

The simple family of complex vector fields $\boldsymbol{E}(\boldsymbol{r}, a, b)$ (3.1), depending on two real parameters $a, b$, illustrates chiral connectedness. Along paths in the $a, b$ plane, $\boldsymbol{E}$ can transform smoothly into its mirror image, without ever being achiral, provided the path never includes the origin $a = b = 0$.

Our analysis has involved the local spin and orbital angular momenta. These are experimentally accessible quantities; they can be separately manipulated and detected by the rotation of small particles in the field [23-27].

The subject of chirality is much wider than the chiral connectedness and application to complex vector fields considered here. Sometimes, chiral connectedness is irrelevant. For example, handedness can be uniquely assigned to knots because the chiral connectedness transformation between mirror images requires strands of the knot to cross: a physically forbidden process.

Another context in which chiral connectedness plays at most a secondary role is the scale-dependence of chirality. People can be left- or right-handed, but their DNA all spirals the same way. It is easy to create geometrical models (or physical LEGO constructions [28]) with a hierarchy of coiled coils, with different handedness at different levels. In observations of such structures, the handedness indicated by any chiral measure will depend on the scale of interrogation (for example, the wavelength of polarised light).

**Acknowledgment**. MVB and KYB thank ICOAM2026 (Graz) for hospitality where this work started. KYB's research has been co-funded by the European Union through the project HORIZON-MSCA-2022-COFUND-01-SmartBRAIN3-101126600.